The Slodderwetenschap (Sloppy Science) of Stochastic Parrots – A Plea for Science to NOT take the Route Advocated by Gebru and Bender

By Michael Lissack (Michael.lissack@isce.edu 617-710-9565)

Tongji University College of Design and Innovation, Shanghai

PREPRINT for ArXIV

Abstract:  This article is a position paper written in reaction to the now-infamous paper titled "On the Dangers of Stochastic Parrots: Can Language Models Be Too Big?" by Timnit Gebru, Emily Bender, and others who were, as of the date of this writing, still unnamed.  I find the ethics of the Parrot Paper lacking, and in that lack, I worry about the direction in which computer science, machine learning, and artificial intelligence are heading.  At best, I would describe the argumentation and evidentiary practices embodied in the Parrot Paper as Slodderwetenschap (Dutch for Sloppy Science) – a word which the academic world last widely used in conjunction with the Diederik Stapel affair in psychology [2].  What is missing in the Parrot Paper are three critical elements: 1) acknowledgment that it is a position paper/advocacy piece rather than research, 2) explicit articulation of the critical presuppositions, and 3) explicit consideration of cost/benefit trade-offs rather than a mere recitation of potential "harms" as if benefits did not matter.  To leave out these three elements is not good practice for either science or research.

**Introduction**

The work of what is referred to as the "Ethical AI" group at Google was brought into some prominence in the public sphere due to events concerning the lead Google author of a paper titled "On the Dangers of Stochastic Parrots: Can Language Models Be Too Big?" (hereafter the "Parrot Paper").  That author, Timnit Gebru, went public at the beginning of December 2020 with complaints about censorship, harm to minorities, systemic racism, and her resignation/termination from Google. [3] The intent of this paper is to deal with the content of that paper (or at least with the content as available from the pre-print which was circulating during the last week or so of December 2020 [1].   This article is NOT about the events surrounding the development of the Parrot Paper but rather is focused on the content of the Parrot Paper itself.  That content seems, at minimum, to meet the definition of Slodderwetenschap (Dutch for Sloppy Science). I believe that the content of the Parrot Paper and the manner by which it is presented raise serious ethical questions. The act of writing a paper raising such questions seems directly contrary to "Ethical AI" – yet the authors write as if they speak for "Ethical AI," an inference readers would have likely drawn had their Google Ethical AI affiliations been allowed to remain attached thereto. [4]

An outside perspective reveals important issues within the Parrot Paper. I am not a computer scientist.  I am a practicing philosopher of science; have led institutions involved with complexity science (from the qualitative perspective) and cybernetics;  ran the ethics program at the Rotterdam School of Management; was editor in chief of a successful academic journal, recently finished editing the second edition of *The Collected Works of Warren McCulloch* [5]; and have developed an alternative search approach based on the idea of "find more like this," which can be found embodied in several tools on the Internet [6]. My academic work includes publications on ethics, management, philosophy of science, complex systems, design, cognitive science, and cybernetics. On two other occasions [7,8], I have called out other examples of Slodderwetenschap, and each time I do so, I hope it is to be the last. That hope remains with this article.

Three disclaimers are needed upfront. First, I have publicly been on record describing the actions of Dr. Gebru to be other than those one might rightly expect from a corporate employee who intended to keep their job; I do believe her separation from Google (be it resignation or termination) was appropriate. This article is not about Dr. Gebru's *actions*, but I acknowledge that some readers may insist on drawing a linkage, which I reject. Second, ethics, in my view, is about surfacing the questions which need to be asked in order to judge actions, or hopefully pre-judge them, and thus engage in a self-reflective process where the others who may be party to that action or its consequences are treated with respect and as "Thous" in a Martin Buber "I-Thou" relation [9]. To me, ethics is NOT some set of rules. Ethics, to me, is about process and context. Readers who define ethics as rule-based may need to be reminded of this process focus as they read on. Finally, I am not affiliated with Google or Alphabet in any way other than as a retail customer.

**What is Good About the Parrot Paper**

The very topic of the Parrot Paper is an ethics question: does the current focus on "language models" of an ever-increasing size in the AI/NLP community need a grounding against potential questions of harm, unintended consequences, and "is bigger really better?" The authors thereby raise important issues that the community itself might use as a basis for self-examination. To the extent that the authors of the Parrot Paper succeed in getting the community to pay more attention to these issues, they will be performing a public service.

In its title, the Parrot Paper raises several important issues: (1) what are the risk/benefit trade-offs of using models that are stochastic in nature and which seemingly "parrot" human language use, (2) is there adequate disclosure/discussion to the general public and the media regarding both the stochastic nature of the models and the "parroting" nature of their output and (3) is the use of ever bigger models capable of providing even more statistically-derived correlations the "best" direction to further development of machine intelligence and natural language processing. There exist alternatives to ever-larger models and to pure stochastic-processing. How might the community weigh the trade-offs involved in determining the appropriate model to employ in a given circumstance?

The Parrot Paper correctly identifies an "elephant in the room" for the MI/ML/AI/NLP community: the very basis by which these large language models are created and implemented can be seen as multi-layer neural network-based black boxes – the input is observable, the programming algorithm readable, the output observable, but HOW the algorithm inside that black box produces the output is not articulable in terms humans can comprehend. [10] What we know is some form of "it works." The Parrot Paper authors prompt readers to examine what is meant by "it works." Again, a valuable public service is being performed by surfacing that question.

The Parrot Paper questions the community's emphasis on leaderboards and asks if, instead, there might not be other potentially "better" ways to measure "improvements" and "success." The authors suggest a series of reporting measures that could be useful in reviewing research and outputs in the community. They ask if the current manner by which success is measured leads the research community AWAY from efforts that might "facilitate long-term progress towards natural language understanding." [PP line 123]

Most importantly, in my view, the Parrot Paper authors remind readers that potential harm lies in both the careless use/abuse of these language models and in the manner by which the outputs of those models are presented to and perceived by the general public. They quote Prabhu and Birhane echoing Ruha Benjamin: "Feeding AI systems on the world's beauty, ugliness, and cruelty, but expecting it to reflect only the beauty is a fantasy." [PP lines 565-567, 11, 12]  The danger they cite is quite real.  When "users" are unaware of the limitations of the models and their outputs, it is all too easy to confuse seeming coherence and exactness for verisimilitude. Indeed, Dr. Gebru first came to public attention highlighting similar dangers with respect to facial recognition software (a danger which remains, unfortunately, with us [13, 14].

The misuse of large language models is only likely to increase, absent the kind of ethical questioning which the authors of the Parrot Paper and others advocate.  What is not stated in the Parrot Paper, but is implicit in some of the questions it raises, is the notion that insufficient research is being devoted to "improving" the incorrect results which emerge a significant percentage of the time from these models. "Success" ratios (the output is deemed to "work") in the sixty percent range and above may seem meaningful to academic researchers, but their implicit failure ratio creates a significant possibility for harm -- which most research projects simply ignore.

It is all well and good that GPT-3 can write what looks like English paragraphs, but to allow those paragraphs to be passed off as "coherent thought" is dangerous and potentially harmful.  The product of a statistical multi-layered neural network vector analysis is not a product of coherent reasoning, no matter how much it may resemble the same.  The public needs to be continually warned about the risks of this mismatch.

If the Parrot Paper were written in a manner that made the above arguments clear and supported them with evidence – both pro and con – the paper would be an important contribution to knowledge.

Alas, that is not the road the authors chose.

**Argumentation Form**

The Parrot Paper is presented as if it were a research paper – that is, as the product of research on a topic with a review of the relevant literature.  The 128 references at the end and the nearly continuous use of citations throughout (with minimal use of quotations) serve to strengthen that presentation. The Parrot Paper is a one-sided position paper with next to no referencing of alternative points of view. Sadly, there is no acknowledgment by the authors anywhere in the paper's text that they are engaged in advocating a set of beliefs.  Advocacy disguised as research represents a similar kind of harm as presenting the output of GPT-3 as coherent reasoning.  The general public is unlikely to be aware of the distinction, and the media even more likely to further the misrepresentation in its reporting.

Researchers and scientists are, of course, free to hold opinions and beliefs.  They are free to argue in advocacy of those positions.  But, when researchers and scientists do so in a manner that the general public and the media are likely to believe is "research," they engage in what borders on fraud.  Diederik Stapel believed that real data was "too messy," so he made up his data sets so as to present a convincing story to the public [15]. The research behind an advocacy position is similarly fraudulent when only one side of an issue (which may have many sides) is presented, and so too when the context from within which the value judgments are being drawn is not given appropriate notice and flagging. The need to present multiple points of view is "messy," but science is not helped when that necessary step is omitted. The Parrot Paper is riddled with such omissions.

From an argumentation perspective, the Parrot Paper reads as if it is a work of abductive reasoning (explore to find the best inference) – the typical method used in scientific research. The form is, however, but pretense. True abductive reasoning requires that alternative perspectives, hypotheses, and explanations be considered, examined, and then narrowed. The Parrot Paper neglects these steps. It starts from a set of perspectives that need to be articulated in order to be made clear, looks at only those hypotheses which are consistent with the authors' presuppositions, and then offers either one explanation or one plus a nominal straw dog (for example, gratuitous mention of the worst fire season in Australia). All of this technique is fine for taking one side of a debate – but not acceptable when one fails to inform the reader that there even is a debate, much less that there exist multiple perspectives.

Perspectives and assumptions matter. The legitimacy of the Parrot Paper's main arguments rests on a set of somewhat hidden but nonetheless critical presuppositions. These critical presuppositions may not be shared by all of the Parrot Paper's readers and are unlikely to be broadly shared by the general public. In this case, because members of the media share the Parrot Paper authors' presuppositions, the risk of the work being presented as "fact" when it is really "opinion" is great. The human tendency for confirmation bias is thus present in how the Parrot Paper is presented. Those members of the general public who do NOT share in the presuppositions are given yet another reason to "distrust" science. For them, the confirmation bias will run in the opposite direction – namely, that the supposed "science" in the Parrot Paper is nonsense.

The Parrot Paper is not science – it is opinion, and science suffers from the authors' deliberate mislabeling.

**The Critical Presuppositions**

Presuppositions are a set of assumptions and beliefs used by a cognizer (one who is engaged in arriving at an understanding) to filter and frame the context and evidence being observed as the initial step in sense-making. These presuppositions come prior to the act of attending to context, observations, or evidence and act as a set of constraints regarding the narrative which a cognizer puts forth to establish an understanding or an explanation. Lissack [16,17] argues why the careful explication of presuppositions is important if science, in general, is to avoid suffering negative effects from the confirmation biases of a less than knowledgeable public. Only if presuppositions are articulated and made explicit can a third party begin to probe what it is about evidence, context, and affordances that allows some label, category, or short narrative to serve as a basis for meaning and understanding.

For clarity, I present what *I see* as the critical presuppositions of the Parrot Paper. I believe they correctly identify the intellectual framing and mindset necessary to understand how the authors of the Parrot Paper came to write it. The presuppositions listed below justify the opinions and beliefs expressed in the Parrot Paper.

1. Ethics and fairness are to be assessed from a Rawlsian [18] perspective. Actions that may result in benefits to those who are not the least well off (or otherwise describable as being the "most marginal") are to be avoided. Only actions that also provide benefits to the most marginal are deemed ethical and fair. [PP lines 94-98, 327, 389-407, 1021-22]
2. Judgments regarding presupposition #1 are to be made based upon first-order effects (the direct result of actions). Making use of second (indirect consequences and entailments) and higher-order effects in argumentation is "suspect." [PP lines 95-97, 322-350]
3. Greater deployment potential for tools means a larger sphere of influence for the model on which the tools depend [PP line 266, Section 4.1]
4. Environmental costs and impacts cannot *ever* be offset and are meaningful first-order effects that must always be minimized. [PP Lines 94-100, Section 3]
5. Systemic racism must be recognized as present in all systems examined and is required to be addressed by any actions taken. [PP lines 385-386, 427-428]
6. Microaggressions and the possibility of encountering upsetting or derogatory language is a significant harm that must either not be allowed to occur or which must be offset. [PP lines 771-789, 813-827]
7. Resources need to be allocated equitably (with no definition given of what counts as equitable or who gets to determine judgments of equality). [PP lines 42-428, 883-887]
8. The dominance of English is a problem (even in an English language setting). [PP lines 18, 327-333, 384, 766-770]
9. Language use, in general, reflects White supremacist, misogynistic, and ageist views. These views must be countered or eliminated. [PP lines 385-386, 822-827]
10. Black-box methods (where only the input and the output are observable while the "how" is unexplained) are, by definition, inferior to articulatable methodologies under ALL circumstances. [PP Lines 119-124, Section 5]
11. Curation – the deliberate and considered selection of the specific materials to be looked at by observers and, in this case, the language models, -- can prevent bias by preventing attention to and consideration of materials not selected to be in the curated set. [PP lines 537- 543, Section 4.4]
12. There exist thought and knowledge leaders who can do the curating called for by presupposition #11 while taking into account the remaining presuppositions. [PP Section 4.4 ]

If this list of twelve presuppositions has a significantly large overlap with your world-view, then the arguments presented in the Parrot Paper are seemingly logical consequences. If, by contrast, you can only agree with some items on the list, the need for a discussion of trade-offs becomes apparent. The extent to which you agree with presuppositions #1 and #2, for example, will then affect how you might go about assessing the requisite trade-offs.

The Parrot Paper would be much stronger if it explicitly acknowledged these presuppositions upfront and then stated that it was a position paper derived from such a world view. No one would then make the error of thinking it was a science paper or a research work. Instead, a reader is left to discover implicit references to each of these presuppositions (as I have highlighted above). By inference, science is left holding the bag of "responsibility" for "explaining" what the Parrot Paper authors leave out – contrasting evidence, a discussion of theory, and a basis for prediction. Those who even partially reject some of the presuppositions are left doubting the scientific practices which supposedly shaped the research, even though the work is devoid of both scientific practice and research.

The Parrot Paper, for example, complains about the environmental impact of the large language models and supplies a semblance of exactness with the inclusion of a related table outlining supposed environmental costs. [PP lines 233-244]  But the Parrot Paper fails to discuss how often and under what circumstances these costs are incurred and what benefits result from each such run of the models.  Without both benefits and costs disclosed, a reader has no basis for making any evaluation regarding trade-offs.  Instead, the Parrot Paper authors rely on the presupposition that any incurrence of cost is bad.  The Parrot Paper is also devoid of any discussion of efforts to improve the efficiency of the models employed, the chips on which they run, and the machines themselves – all of which will reduce environmental impacts.  Huge environmental improvements are forecast for the near future, but that "fact" is missing from the Parrot Paper's discussion of environmental issues. [19]

The hidden complaint in the Parrot Paper is that the "real world" contains both ambiguity and "bad language." [PP lines 802-809] The authors of the Parrot Paper believe the world would be better off if the ambiguity could be eliminated (or at least mitigated) and the same with the use of "bad language."  Large language models succeed in statistically capturing and then replaying ("parroting") both the bad language and the ambiguity.  By contrast, a smaller, more ordered, and curated set of models could eliminate the "bad" form of the language and inherent ambiguities in language use. [PP lines 994-999]

If one is building a tool with a specific set of use cases which are known in advance, curation of the dataset input to train the model used in such a tool can produce both efficiencies and "better results." Having built a research-oriented search tool based on the concept of curation [20], I recognize the appeal of such an ordered approach.  But the assertion of order comes at a cost: there is little resilience (the ability to deal with the unexpected), weak signals have a hard time getting noticed, and the sagacity required for serendipity is actively discouraged. [21]  The Parrot Paper authors believe that their curation-approach can overcome these limitations.   Yet, their paper is devoid of any explanation of how.

Curation can also pose its own dangers – harms not discussed in the Parrot Paper – involving censorship and forbidden language.  If a curated AI model is deployed to monitor discussions amongst scientists and researchers, what is to stop that model from preventing discussion of words deemed "unacceptable" by the curators? [22]  Is science (and indeed free speech) well-served if the curation desired by Gebru and Bender is allowed to somehow oversee discourse?  Galileo Galilei suffered from such curation.  Science was not better off.

The Parrot Paper, in short, is a plea for more control, more order, to be guided by the "wise." Who is "wise" is not defined.

**Why the Parrot Paper Approach is Bad for Science**

Science and research, as expressed in the Parrot Paper, are model-based activities.  Yet, the concept of model expressed therein leaves little room for multiple interpretations, context-driven choices, and hypothetical "what-if" interventions. Model, as used in the Parrot Paper, is a static representation of a situation, context, or even the world. The field of Cybernetics sees models quite differently.  To the cybernetician, it is the very ability of an actor to question interpretations and to make choices (what philosophers call agency) that distinguishes a model from a static representation [23, 24, 25]. To be a model (and not just a static representation), the user of the model must have agency. The science of the Parrot Paper, on the other hand, is an engineered, curated work of static representations.

Because the Parrot Paper authors wish an escape from ambiguity and bad language, actors are denied agency. Agency is problematic to the Parrot Paper authors because it allows each user of a model to choose from among multiple interpretations, individual histories, and context-dependence – choices which the Parrot Paper authors assert hold the potential to "harm." Agency is what creates the opportunity for the new, the emergent, and the innovative. Agency allows for progress and growth. Most scientists operate under the impression that their work will expand the opportunities and resources available to the world. By contrast, the Parrot Paper seems to assume a zero-sum world where the innovative, emergent, and new represent further threats to the marginalized underclass. In such a world, given the presuppositions outlined above, it is imperative that efforts be made to redistribute some of the resources and efforts from the non-marginalized to the marginalized. With an "expanding pie," gains may accrue to the collective "all" but not equally, probably not "fairly," and surely not on the same timelines. Such gains defy a Rawlsian concept of fairness and justice. Remember, to Rawls, "just" gains are those which provide at least some benefit(s) to the most marginalized. These benefits are required to be both first-order effects and measurable. Given the Parrot Paper authors" assertion of Rawls, they reject the potential second and higher-order benefits of prospectively "non-just" growth and instead assert that the scientific efforts in AI/ML/MI/NLP be directed to zero-sum redistribution.

Science cannot advance using the zero-sum approach.

The Parrot Paper authors argue for curation and control. [PP Section 4.4] They seek to ban the black box in favor of the transparent. [PP Section 4.3] They give priority to first-order easily observed effects and see second-order and higher effects as potentially harmful. This is an engineering-based approach. It assumes problems are well defined. But, we do not live in such a well-defined world. Our world is messy, complex, and chaotic. Every choice involves one or more trade-offs.

One goal of science is exploring the chaotic and unraveling the mysteries of the complex. The Parrot Paper authors demonstrably prefer a world of clarity and simplicity. Indeed, their disdain for the "unfathomable" elements of large language models [PP line 124] is captured in a rejection of complexity itself. Note there is a distinction between the com*plic*ated and the com*plex*. With the complicated (the Latin stem *plic* means to fold), one can "unfold" a problem or situation to expose a simple surface potentially sufficient to capture the essence of the issue at hand. By contrast, with the complex (the Latin stem *plex* means to weave), the very attempt to reduce an interwoven situation to a simple surface carries with it the risk of destroying both the problem and its context. (Woven things once unwoven are hard to put back together. Unfolded things can be easily refolded.) The Parrot Paper authors are seemingly seeking to reduce the complexity of the "real world" through the complicated task of curation in a driving quest for a "clearness" over which they can assert control.

Science cannot afford such a quest for control. While a controlled science composed of rigorous models may result in successful predictions, those predictions seldom work outside the narrow sphere or domain in which assumed ceteris paribus constraints hold (all other items are assumed to remain constant). Within a controlled situation, scientists may assert that their results sufficiently describe the "essence" of the issue at hand, such that ambiguities do not matter. However, the general public is seldom willing to ascribe confidence in such an assertion. They will either accept the model as "reality" or reject the very science which produced it. [26]

In complex environments, the key to achieving innovation, growth, and resilience is to partially embrace complexity itself and to use it to ask questions, probe sensitivities, examine dependencies, and seek to determine what presuppositions are needed for a given meaning to hold. It is the role of science and of scientific researchers to then ask what happens should the values assumed for those presuppositions change. I will go further: ***only by engaging in such questioning can one be assured that the choices one makes with regard to trade-offs are ethical***.

The questions which practicing scientists need to ask are themselves quite simple: Are we looking at the right things in our environment?

- Do we have the right words, models, or representations to express what we are thinking?
- Are we sufficiently self-aware of our beliefs?
- Do we know the limitations of what it is we think we know?

A science whose practice is to ask such questions differs from the top-down curation techniques advocated by the Parrot Paper authors. The Parrot Paper authors implicitly assert that they (and others "like them") know how to manage whatever trade-offs may arise between costs and benefits. Science cannot afford such hubris.

Those who research and practice in the fields of machine intelligence, computer science, and NLP take note: Progress, innovation, and the next set of breakthroughs arise from asking questions and generating new answers. To embark down the road of curation based on present knowledge is a dead end. An end that must be avoided.

The goals of the Parrot Paper seem noble, but its execution is ethically flawed. The methodologies and priorities called for by the Parrot Paper's authors cannot be supported by practicing scientists. Science is not advanced by demanding top-down control, curation, and reductionism. It cannot progress if the possibility of emergence is seen as threatening. Taking such a reductionist view is Slodderwetenschap. We cannot afford to be sloppy.

**References**

[1]     Gebru, T. et al (2021) "On the Dangers of Stochastic Parrots: Can Language Models Be Too Big?" In FAccT '21: ACM FAccT Conference 2021, March 2021, Online. ACM, New York, NY, USA, 12 pages. https://doi.org/10.1145/1122445.1122456 as retrieved from https://gofile.io/d/WfcxoF on 12/25/20

[2]     Bhattacharjee, Y. (2013) "The mind of a Con Man" NY Times, April 26 https://www.nytimes.com/2013/04/28/magazine/diederik-stapels-audacious-academic-fraud.html

[3]     Tiku, N. (2020) "Google hired Timnit Gebru to be an outspoken critic of unethical AI. Then she was fired for it.", Washington Post, December 23 https://www.washingtonpost.com/technology/2020/12/23/google-timnit-gebru-ai-ethics/

[4]     Allyn, B. (2020) "Google AI Team Demands Ousted Black Researcher Be Rehired And Promoted" NPR, December 17, https://www.npr.org/2020/12/17/947413170/google-ai-team-demands-ousted-black-researcher-be-rehired-and-promoted

[5]     McCulloch, W. (2021) The Collected Works of Warren S. McCullocg, Second Edition, edited by Lissack, M and Holland, A., Emergent Publications and The American Society for Cybernetics, https://eco.emergentpublications.com/McCulloch

[6]     http://myresearchtool.com and http://findrelatedbooks.com

[7]     Lissack, M. and Richardson, K. (2001) "When Modeling Social Systems, Models ≠ the Modeled: Reacting to Wolfram's A New Kind of Science", Emergence, Vol 3, Issue 4, 95-111, https://www.tandfonline.com/doi/abs/10.1207/S15327000EM0304_7

[8]     Lissack, M. (2013) "Subliminal Influence Or Plagiarism By Negligence?: The Slodderwetenschap Of Ignoring The Internet" The Journal of Academic Ethics, December https://www.researchgate.net/publication/259065229

[9]     Buber, Martin, (1923,2000) *I and Thou*, Scribner, New York

[10]    Mittelstadt, B.D. et al, (2016)" The ethics of algorithms: Mapping the debate", *Big Data & Society*, December https://journals.sagepub.com/doi/10.1177/2053951716679679

[11]    Prabhu, V. and Birhane, A. (2020) "Large image datasets: A pyrrhic win for computer vision?" https://arxiv.org/abs/2006.16923

[12]    Ruha Benjamin. 2019. *Race After Technology: Abolitionist Tools for the New Jim Code*. Polity Press, Cambridge, UK

[13]    Najibi, A (2020) "Racial Discrimination in Face Recognition Technology" http://sitn.hms.harvard.edu/flash/2020/racial-discrimination-in-face-recognition-technology/

[14]    Gebru, T. (2020) "Race and Gender" *The Oxford Handbook of Ethics of AI* edited by Dubber, M et al at https://www.oxfordhandbooks.com/view/10.1093/oxfordhb/9780190067397.001.0001/oxfordhb-9780190067397-e-16 and https://arxiv.org/ftp/arxiv/papers/1908/1908.06165.pdf

[15]    Stapel, D. (2016) *Faking Science: A True Story of Academic Fraud* translated by Brown, N. http://nick.brown.free.fr/stapel/FakingScience-20161115.pdf

[16]    Lissack, M. (2017) "Second Order Science: Examining Hidden Presuppositions in the Practice of Science," *Foundations of Science*, 22, 557–573 https://link.springer.com/article/10.1007/s10699-016-9483-x

[17]    Lissack, M. (2017) "What Second Order Science Reveals About Scientific Claims: Incommensurability, Doubt, and a Lack of Explication," *Foundations of Science*, 22, 575–593 https://link.springer.com/article/10.1007/s10699-016-9484-9

[18]    Rawls, John (1971) *A Theory Of Justice*, Belknap Press, Cambridge, MA.

[19]    Imani, M. et al (2020) "DUAL: Acceleration of Clustering Algorithms using Digital-based Processing In-Memory," 53rd Annual IEEE/ACM International Symposium on Microarchitecture (MICRO),

Athens, Greece, pp. 356-371, doi: 10.1109/MICRO50266.2020.00039. https://ieeexplore.ieee.org/document/9251944

[20]     http://myresearchtool.com

[21]     Lederach, J. (2005) "On Serendipity: The Gift of Accidental Sagacity" in *The Moral Imagination: The Art and Soul of Building Peace,* Oxford University Press, NY, https://oxford.universitypressscholarship.com/view/10.1093/0195174542.001.0001/acprof-9780195174540-chapter-11

[22]     Maroja, L (2019) "Self-Censorship on Campus Is Bad for Science" The Atlantic, May 28 https://www.theatlantic.com/ideas/archive/2019/05/self-censorship-campus-bad-science/589969/

[23]     Lissack, M. (2019) "Understanding Is a Design Problem: Cognizing from a Designerly Thinking Perspective." Part 1, *She Ji: The Journal of Design, Economics, & Innovation,* Volume 5, Issue 3, Pp. 231-246,  https://www.sciencedirect.com/science/article/pii/S2405872619300589

[24]     Lissack, M. (2019) "Understanding Is a Design Problem: Cognizing from a Designerly Thinking Perspective." Part 1, *She Ji: The Journal of Design, Economics, & Innovation,* Volume 5, Issue 4, Pp. 327-342,  https://www.sciencedirect.com/science/article/pii/S2405872619300929

[25]     Lissack, M. (2011) "When explanations "cause" error: A look at representations and compressions," Information Knowledge Systems Management, vol. 10, no. 1-4, pp. 189-212, https://content.iospress.com/articles/information-knowledge-systems-management/iks00193

[26]     Funk, C. et al. (2019) "Trust and Mistrust in Americans' Views of Scientific Experts" https://www.pewresearch.org/science/wp-content/uploads/sites/16/2019/08/PS_08.02.19_trust.in_.scientists_FULLREPORT_8.5.19.pdf
2